\documentclass[floatfix,showpacs]{revtex4}
\usepackage{epsfig}
\usepackage{graphicx}
\newcommand{\vphi}{\varphi}

\begin{document}

\title{Gravitating Stationary Dyons and Rotating Vortex Rings}

\author{Burkhard Kleihaus, Jutta Kunz, and Ulrike Neemann}
\affiliation{Institut f\"ur Physik, Universit\"at Oldenburg,
D-26111, Oldenburg, Germany}

\date{\today}

\begin{abstract}

We construct dyons, and electrically charged
monopole-antimonopole pairs and vortex rings 
in Yang-Mills-Higgs theory coupled to Einstein gravity.
The solutions are stationary, axially symmetric and asymptotically flat.
The dyons with magnetic charge $n\ge 2$ represent non-static solutions
with vanishing angular momentum.
The electrically charged monopole-antimonopole pairs and vortex rings,
in contrast, possess vanishing magnetic charge, but finite angular momentum,
equaling $n$ times their electric charge.
\end{abstract}
\maketitle

\section{Introduction}

The non-trivial vacuum structure of SU(2) Yang-Mills-Higgs (YMH) theory 
gives rise to regular non-perturbative finite energy solutions, 
such as magnetic monopoles, multimonopoles and monopole-antimonopole systems.
While spherically symmetric monopoles carry unit topological
charge \cite{mono,jul,PraSom}, monopoles with charge $n>1$
\cite{WG,RR,mmono,KKT}
and monopole-antimonopole systems \cite{Taubes,Rueber,mapKK,KKS}
are axially symmetric 
or possess no rotational symmetry at all \cite{CorGod,monoDS}.

To any static solution of the YMH equations
there corresponds a family of electrically charged solutions 
\cite{jul,PraSom,wein,dyonhkk}.
In the Prasad-Sommerfield limit these electrically charged solutions
are obtained directly from the electrically neutral solutions
via simple scaling relations,
by requiring the time component of the gauge field and the Higgs field
to be parallel \cite{wein,dyonhkk}.

Bogomol'nyi-Prasad-Sommerfield (BPS) monopoles and
Julia-Zee dyons do not admit slowly rotating excitations \cite{HSV}.
Indeed, monopoles and dyons cannot rotate
in the sense, that they possess finite angular momentum,
since they are globally regular solutions 
which carry magnetic charge \cite{radu}.
Thus monopoles and dyons with higher magnetic charge cannot rotate either.
Whereas BPS dyons also have vanishing angular momentum density,
non-BPS dyons with higher magnetic charge might possess
a finite angular momentum density,
yielding a vanishing angular momentum upon integration, though,
because of symmetry reasons.

In monopole-antimonopole pairs, on the other hand, the magnetic charge vanishes
\cite{Taubes,Rueber,mapKK,KKS}.
In these axially symmetric solutions the two nodes of the Higgs field, 
representing the location of the magnetic charges,
are situated symmetrically on the positive and negative $z$-axis.
When electric charge is added to both the monopole and antimonopole in the pair,
they experience a repulsive force and the poles move further apart \cite{dyonhkk}.
More importantly, however, the pair begins to rotate about its symmetry axis,
yielding an angular momentum equal to its electric charge, $J=Q$ \cite{radu,PRT}.
Thus the presence of electric charge 
again leads to a finite angular momentum density.
But since the magnetic charge of the pair vanishes,
it may rotate, and indeed it must rotate with $J=Q$ \cite{radu,PRT}.

Mono\-pole--antimonopole pairs
can also be formed from doubly charged
monopoles and antimonopoles ($n=2$) \cite{tigran,KKS}.
For higher values of $n$, in constrast,
a completely different type of solution appears \cite{KKS}.
Here the Higgs field vanishes on a ring centered around the symmetry axis.
Therefore we refer to these solutions as vortex rings.
Adding electric charge to these magnetically neutral solutions
then should yield 
rotating $n=2$ mono\-pole--antimonopole pairs
and rotating vortex rings possessing angular momentum $J=nQ$.

When gravity is coupled to YMH theory,
gravitating monopoles \cite{gmono,HKK},
gravitating monopole-antimonopole pairs
and gravitating vortex rings arise \cite{MAP,KKS4}.
In each case, a branch of gravitating solutions
emerges smoothly from the corresponding flat space solution,
and extends up to a maximal value of the coupling constant,
where, for vanishing Higgs self-coupling constant,
it merges with a second branch \cite{foot}.
For monopoles this second branch extends only slightly backwards
before it merges with the branch of extremal Reissner-Nordstr\"om black holes
\cite{gmono,HKK,foot0}.
For monopole-antimonopole pair solutions and vortex rings,
in contrast,
this second branch extends all the way back to vanishing coupling constant,
where the solutions shrink to zero size.

The coupling constant $\alpha$,
entering the Einstein-Yang-Mills-Higgs (EYMH) equations,
is proportional to the Higgs vacuum expectation value $v$
and the square root of the gravitational constant $G$.
Variation of $\alpha$ may thus be considered
as variation of the gravitational constant $G$ along the first branch
and as variation of the Higgs vacuum expectation value $v$
along the second branch.
Consequently, the Higgs field vanishes
in the limit $\alpha \rightarrow 0$ on the second branch,
and, after scaling the coordinates, the mass and the Higgs field with $\alpha$,
solutions of Einstein-Yang-Mills (EYM) theory are obtained
\cite{MAP,KKS4},
which correspond to the lowest mass Bartnik-McKinnon (BM) solution \cite{BM} 
or its generalizations \cite{KK,IKKS}.

When dyons \cite{gdyon} or electrically charged
monopole-antimonopole pair solutions \cite{PRT}
are coupled to gravity, analogously
a corresponding branch of gravitating dyons 
or electrically charged monopole-antimonopole pair solutions
emerges smoothly from the respective flat space solution \cite{gdyon,PRT}.
Whereas gravitating dyons again merge with the corresponding
branch of extremal Reissner-Nordstr\"om black holes
at a critical value of the coupling constant \cite{gdyon,foot0},
the critical behaviour for gravitating 
electrically charged monopole-antimonopole pair solutions
could not be resolved previously \cite{PRT}.

Gravitating spherically symmetric dyons are static 
in the sense, that their angular momentum density vanishes.
In this letter we show, that 
axially symmetric gravitating dyons ($n\ge 2)$ are stationary, 
while they carry no angular momentum.
We further resolve the critical behaviour
of electrically charged monopole-antimonopole pair solutions,
and we construct rotating vortex ring solutions.
In all calculations we limit ourselves to vanishing Higgs self-coupling.

In section II we present the action, the axially
symmetric Ansatz and the boundary conditions.
In section III we discuss
the properties of stationary dyons, and 
rotating monopole-antimonopole pairs and vortex rings.
We present our conclusions in section IV.

\section{\bf Einstein-Yang-Mills-Higgs Solutions}

\subsection{Action}

We consider the SU(2) EYMH action
in the limit of vanishing Higgs potential,
\begin{eqnarray}
S &=& \int \left [ \frac{R}{16\pi G}
  - \frac{1}{2} {\rm Tr} \left(F_{\mu\nu} F^{\mu\nu}\right)
 -\frac{1}{4} {\rm Tr} \left( D_\mu \Phi D^\mu \Phi \right)
 \right ] \sqrt{-g}\, d^4x
\ \label{action} \end{eqnarray}
with curvature scalar $R$,
SU(2) field strength tensor
\begin{equation}
F_{\mu \nu} =
\partial_\mu A_\nu -\partial_\nu A_\mu + i e \left[A_\mu , A_\nu \right]
\ , \label{fmn} \end{equation}
gauge potential $A_\mu = 1/2 \tau^a A_\mu^a$,
gauge covariant derivative
\begin{equation}
D_\mu = \nabla_\mu +ie [ A_\mu, \cdot \ ]
\ , \label{Dmu} \end{equation}
and Higgs field $\Phi = \tau^a \Phi^a$;
$G$ is Newton's constant, and $e$ is the gauge coupling constant.
We impose a Higgs field vacuum expectation value $v$.

Variation of the action Eq.~(\ref{action}) with respect to the metric
$g_{\mu\nu}$, the gauge potential $A_\mu^a$, and the Higgs field $\Phi^a$
leads to the Einstein equations and the matter field equations,
respectively.

\subsection{Ans\"atze}

We consider regular stationary, axially symmetric solutions
with Killing vector fields $\xi=\partial_t$ and $\eta=\partial_{\varphi}$.
We employ
the Lewis-Papapetrou form of the metric in isotropic coordinates
\cite{kkrot}
\begin{equation}
ds^2 = -fdt^2+\frac{m}{f}\left[dr^2+r^2 d\theta^2\right] 
       + \frac{l r^2 \sin^2\theta}{f}
          \left[d\varphi-\frac{\omega}{r}dt\right]^2 \  
\ .  \label{am} \end{equation}
The gauge potential is parametrized by \cite{kkrot}
\begin{equation}
A_\mu dx^\mu
  =   \left( B_1 \frac{\tau_r^{(n,m)}}{2e} + B_2 \frac{\tau_\theta^{(n,m)}}{2e} \right) dt
-n\sin\theta\left[H_3 \frac{\tau_r^{(n,m)}}{2e}
            +(1-H_4) \frac{\tau_\theta^{(n,m)}}{2e}\right] d\vphi
+\left(\frac{H_1}{r}dr +(1-H_2)\, d\theta \right)\frac{\tau_\vphi^{(n)}}{2e}
\ , \label{a1} \end{equation}
and the Higgs field by \cite{KKNeymh}
\begin{equation}
\Phi =v \left( \Phi_1 \tau_r^{(n,m)} + \Phi_2 \tau_\theta^{(n,m)} \right)
\ , \label{a2} \end{equation}
where $n$ and $m$ are integers,
with $\pm n$ representing the magnetic charge of single (anti)monopoles
in monopole-antimonopole chains
and $m$ the total number of monopoles and antimonopoles
in monopole-antimonopole chains \cite{KKS}.
Dyons are obtained for $m=1$, 
and monopole-antimonopole pairs and vortex rings for $m=2$.
The $su(2)$ matrices
$\tau_r^{(n,m)}$, $\tau_\theta^{(n,m)}$, and $\tau_\vphi^{(n)}$
are defined as scalar products of the spatial unit vectors
\begin{eqnarray}
{\hat e}_r^{(n,m)} & = & \left(
\sin(m\theta) \cos(n\vphi), \sin(m\theta)\sin(n\vphi), \cos(m\theta)
\right)\ , \nonumber \\
{\hat e}_\theta^{(n,m)} & = & \left(
\cos(m\theta) \cos(n\vphi), \cos(m\theta)\sin(n\vphi), -\sin(m\theta)
\right)\ , \nonumber \\
{\hat e}_\vphi^{(n)} & = & \left( -\sin(n\vphi), \cos(n\vphi), 0 \right)\ ,
\label{unit_e}
\end{eqnarray}
with the Pauli matrices $\tau^a = (\tau_x, \tau_y, \tau_z)$.
All functions depend on $r$ and $\theta$, only.

The ansatz is form-invariant under Abelian gauge transformations $U$
\cite{kkrot}
\begin{equation}
 U= \exp \left({\frac{i}{2} \tau_\varphi^{(n)} \Gamma(r,\theta)} \right)
\ .\label{gauge} \end{equation}
With respect to this residual gauge degree of freedom
we choose the gauge fixing condition
$r\partial_r H_1-\partial_\theta H_2 =0$
\cite{kkrot}.

\subsection{Boundary Conditions}

Regularity of the solutions at the origin ($r=0$) 
requires for the metric functions the boundary conditions 
\begin{equation}
\partial_r f(r,\theta)|_{r=0}= 
\partial_r m(r,\theta)|_{r=0}= 
\partial_r l(r,\theta)|_{r=0}= 0
\ , \label{bc2a} \end{equation}
whereas the gauge and Higgs field functions $H_i$ and $\Phi_i$ satisfy
\begin{equation}
H_1(0,\theta)= H_3(0,\theta)= 0\ , \ \ \ \
H_2(0,\theta)= H_4(0,\theta)= 1 \ ,
\end{equation}
and for even $m$
\begin{equation}
\sin(m\theta) B_1(0,\theta) + \cos(m\theta) B_2(0,\theta) = 0 \ ,
\end{equation}
\begin{equation}
\left.\partial_r\left[\cos(m\theta) B_1(r,\theta)
              - \sin(m\theta) B_2(r,\theta)\right] \right|_{r=0} = 0 \ ,
\end{equation}
\begin{equation}
\sin(m\theta) \Phi_1(0,\theta) + \cos(m\theta) \Phi_2(0,\theta) = 0 \ ,
\end{equation}
\begin{equation}
\left.\partial_r\left[\cos(m\theta) \Phi_1(r,\theta)
              - \sin(m\theta) \Phi_2(r,\theta)\right] \right|_{r=0} = 0 \ ,
\end{equation}
whereas for odd $m$ $B_i(0,\theta)=\Phi_i(0,\theta)=0$.

Asymptotic flatness imposes on the metric functions
at infinity ($r=\infty$) the boundary conditions
\begin{equation}
f \longrightarrow 1 \ , \ \ \ 
m \longrightarrow 1 \ , \ \ \ 
l \longrightarrow 1 \  , \ \ \
\omega \longrightarrow 0 \  . \ \ \
\   \label{bc1a} \end{equation}

The boundary conditions for the functions $H_1 - H_4$, $B_1$, $B_2$, 
$\Phi_1$, $\Phi_2$ read
\begin{equation}
H_1 \longrightarrow 0 \ , \ \ \ \
H_2 \longrightarrow 1 - m \ , \ \ \ \
\label{K12infty}
\end{equation}
\begin{equation}
H_3 \longrightarrow \frac{\cos\theta - \cos(m\theta)}{\sin\theta}
\ \ \ m \ {\rm odd} \ , \ \ \
H_3 \longrightarrow \frac{1 - \cos(m\theta)}{\sin\theta}
\ \ \ m \ {\rm even} \ , \ \ \
\label{K3infty}
\end{equation}
\begin{equation}
H_4 \longrightarrow 1- \frac{\sin(m\theta)}{\sin\theta} \ ,
\label{K4infty}
\end{equation}
\begin{equation}        \label{a0infty}
B_1\longrightarrow  \gamma \ , \ \ \ \ B_2 \longrightarrow 0 \ .
\end{equation}
\begin{equation}        \label{Phiinfty}
\Phi_1\longrightarrow  1 \ , \ \ \ \ \Phi_2 \longrightarrow 0 \ .
\end{equation}

On the symmetry axis, we impose \cite{kkrot}
$\partial_\theta f = \partial_\theta m = \partial_\theta l =
\partial_\theta \omega = 0$,
$H_1=H_3=0$,
$\partial_\theta H_2 =\partial_\theta H_4=0$,
$\partial_\theta B_1 =0$, $B_2=0$,
$\partial_\theta \Phi_1 =0$, $\Phi_2=0$.
Regularity further requires $m(r,\theta)=l(r,\theta)$ and 
$H_2(r,\theta)=H_4(r,\theta)$ on the symmetry axis.

\section{Results}

We here discuss our numerical results for
stationary dyons, and rotating monopole-antimonopole pairs and vortex rings,
and determine the dependence of these solutions on the coupling constant $\alpha$.

\subsection{Numerical procedure}

To construct solutions subject to the above boundary conditions,
we map the infinite interval of the variable $r$
onto the unit interval of the compactified radial variable
$\bar r \in [0:1]$,
$$
\bar r = \frac{r}{1+r}
\ , $$
i.e., the partial derivative with respect to the radial coordinate
changes according to
$
\partial_r \to (1- \bar r)^2\partial_{\bar r}
\ . $
The numerical calculations are then performed with the help of the FIDISOL package
based on the Newton-Raphson iterative procedure \cite{FIDI}.

\subsection{Charges}

Let us introduce dimensionless quantities,
\begin{equation}
x = ev  r \ , \ \ \
\hat B_i = \frac{1}{ev} \, B_i \ , \ \ \
\hat \gamma = \frac{1}{ev} \, \gamma \ . \ \ \
\   \end{equation}
For fixed $n$ and $m$, the equations then
depend only on the dimensionless coupling constant $\alpha$ \cite{gmono}
\begin{equation}
\alpha = \sqrt{4 \pi G} v \ ,
\   \end{equation}
since we restrict to vanishing Higgs potential.

The dimensionless mass $M$ and angular momentum $J$ of the solutions
are obtained from the asymptotic expansion of the metric functions
\begin{equation}
M = \frac{1}{2\alpha^2} \lim_{x \rightarrow \infty} x^2 \partial_x  f
\ , \ \ \
J = \frac{1}{2\alpha^2} \lim_{x \rightarrow \infty} x^2 \omega
\ , \label{MJ} \end{equation}
the dimensionless electric charge $Q$ \cite{KKNeymh} 
and magnetic charge $P$ \cite{KKS} are obtained from
\begin{equation}
Q = -\lim_{x \rightarrow \infty} x \left( \hat B_1-\hat \gamma \right)
\ , \ \ \
P = \frac{n}{2}\left(1 - (-1)^m\right)
\ . \label{QP} \end{equation}
Magnetically charged solutions have vanishing angular momentum, $J=0$ \cite{radu}, 
and vanishing magnetic dipole moment, $\mu_{\rm mag}=0$ \cite{KKS}.
Magnetically neutral solutions
possess a finite dipole moment \cite{KKS},
which can be obtained
from the asymptotic form of the non-Abelian gauge field,
after transforming to a gauge where the Higgs field
is constant at infinity, $\Phi = \tau_z$,
\begin{equation}
A_\mu dx^\mu = -{\mu_{\rm mag}}\frac{\sin^2\theta}{x}\frac{\tau_z}{2} d\vphi \ ,
\label{magmom}
\end{equation}
and they satisfy the relation
\begin{equation}
J=
\frac{n}{2}\left(1 + (-1)^m\right) Q
\ , \label{JQ} \end{equation}
generalizing the previous relations \cite{radu}.
The full asymptotic expansion will be given elsewhere \cite{KKNeymh}.

\boldmath
\subsection{Stationary dyons: $n>1$}
\unboldmath

Gravitating dyons with magnetic charge $n=1$ are spherically symmetric
and static \cite{gdyon}.
Their $\alpha$-dependence is completely analogous to the $\alpha$-dependence
of gravitating monopoles. Thus a branch of gravitating dyons
emerges from the flat space dyon solution and extends up to a maximal
value $\alpha_{\rm max}$. 
There it merges with a short second branch, which bends backwards
and then merges at a critical value $\alpha_{\rm cr}$
with the branch of extremal Reissner-Nordstr\"om solutions \cite{gdyon,foot0}.

Like monopoles with higher magnetic charge \cite{HKK}
also dyons with higher magnetic charge show a similar $\alpha$-dependence:
they merge with the corresponding branch of extremal Reissner-Nordstr\"om solutions
\cite{foot0}.
However, numerical accuracy does not allow to definitely conclude, whether 
a short second branch is present, i.e., whether
$\alpha_{\rm max} \ne \alpha_{\rm cr}$ \cite{HKK}.
We exhibit the scaled mass $\alpha M$ 
and the electric charge $Q$ for dyons with magnetic
charge $n=2$ and $n=3$ in Figs.~1. 

\noindent\parbox{\textwidth}{
\centerline{
(a)\mbox{\epsfysize=5.0cm \epsffile{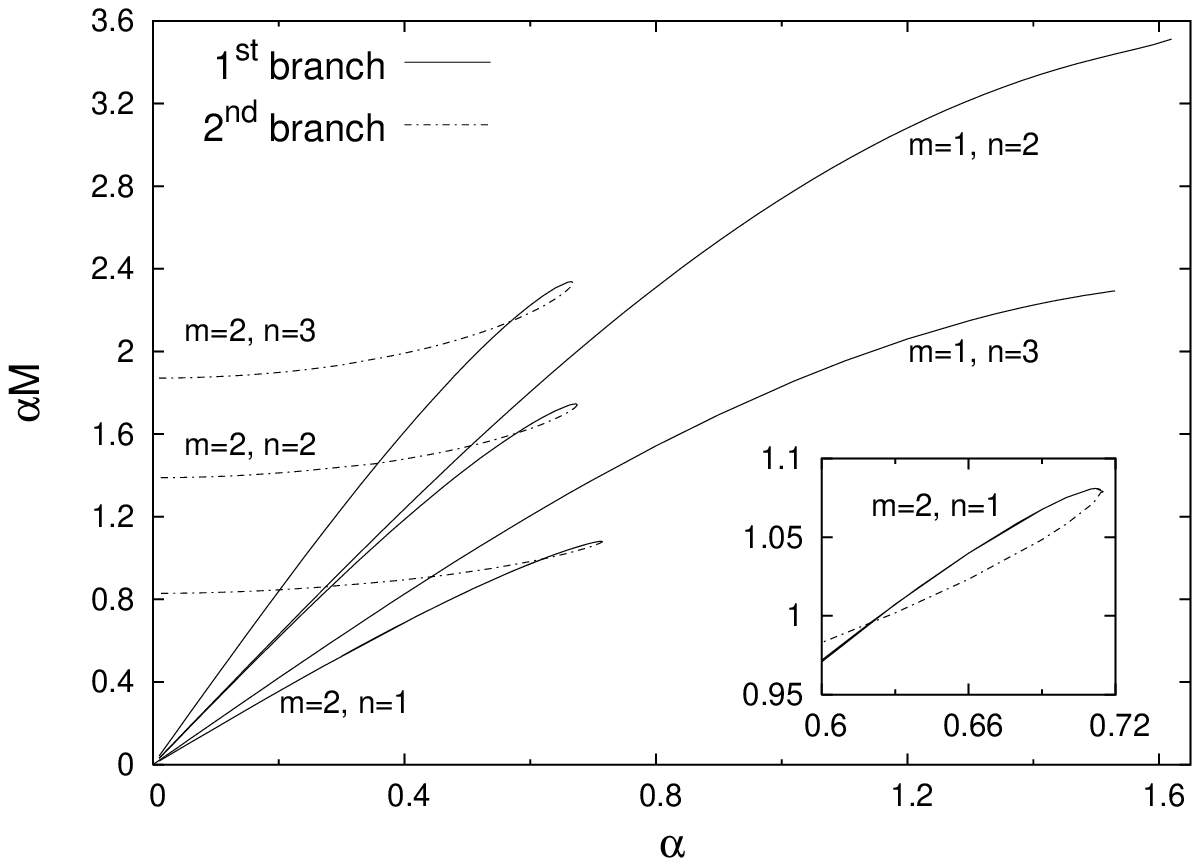} }
\hspace{2cm}
(b)\mbox{\epsfysize=5.0cm \epsffile{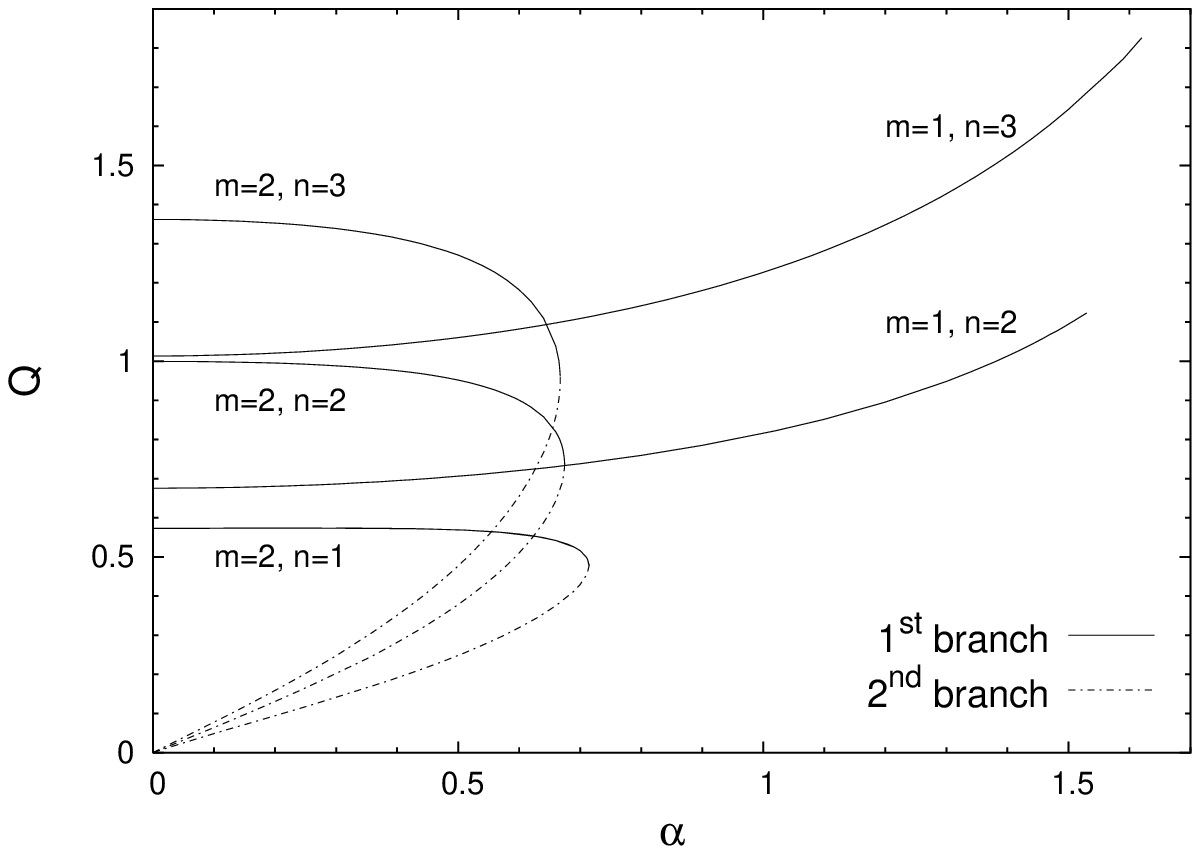} }
}\vspace{0.5cm}
{\bf Fig.~1} \small
The scaled mass $\alpha M$ (a) and the charge $Q$ (b)
are shown as functions of the coupling constant $\alpha$
for dyon solutions with $n=2,3$
and for monopole-antimonopole resp.~vortex ring solutions
with $n=1,2,3$ at $\hat \gamma=0.32$.
\vspace{0.5cm}
}

Gravitating dyons with magnetic charge $n=1$ are static, since they
have vanishing angular momentum density $T^t_{\,\varphi}$.
We here demonstrate that
gravitating dyons with magnetic charge $n>1$ are stationary but not static.
They possess a finite angular momentum density.
But since their angular momentum density is antisymmetric 
with respect to reflection, $z \rightarrow -z$,
their angular momentum vanishes.
In Fig.~2 we exhibit the energy density $T^t_{\,t}$ 
and the angular momentum density $T^t_{\,\varphi}$
for a typical gravitating dyon with $n=2$ for $\alpha=1.4$.
The energy density is toruslike \cite{dyonhkk,HKK}, 
but becomes spherical in the limit $\alpha \rightarrow \alpha_{\rm cr}$.
The angular momentum density vanishes both for $\alpha=0$ and
$\alpha=\alpha_{\rm cr}$.
With increasing $\alpha$ it increases in magnitude,
but is localized in a decreasing region of space.

\noindent\parbox{\textwidth}{
\centerline{
(a)\mbox{\epsfysize=6.0cm \epsffile{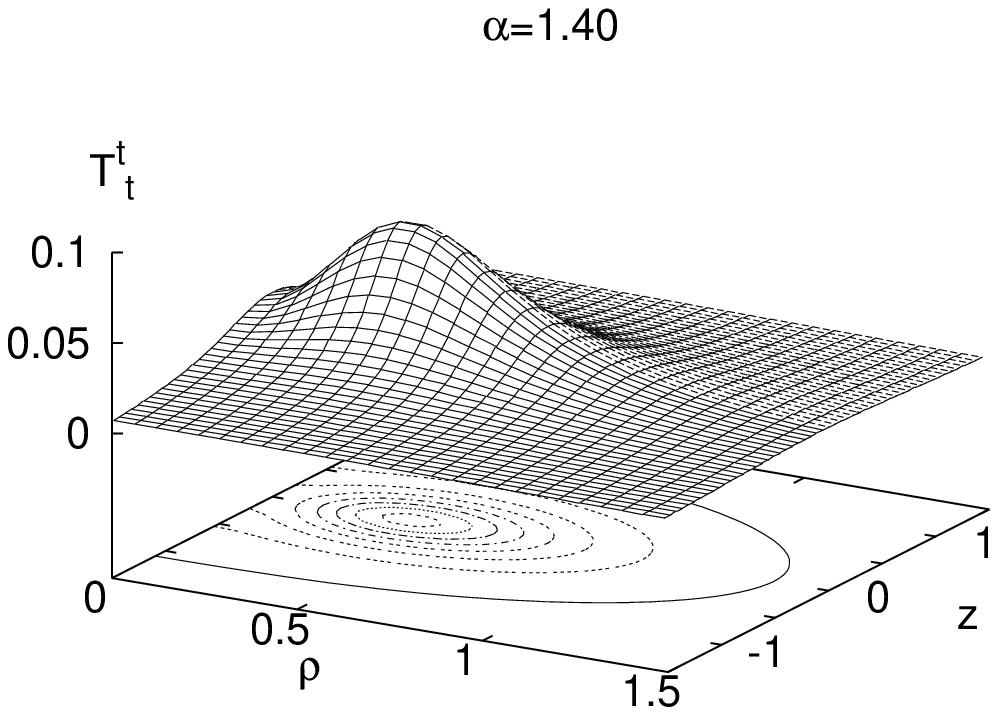} }
(b)\mbox{\epsfysize=6.0cm \epsffile{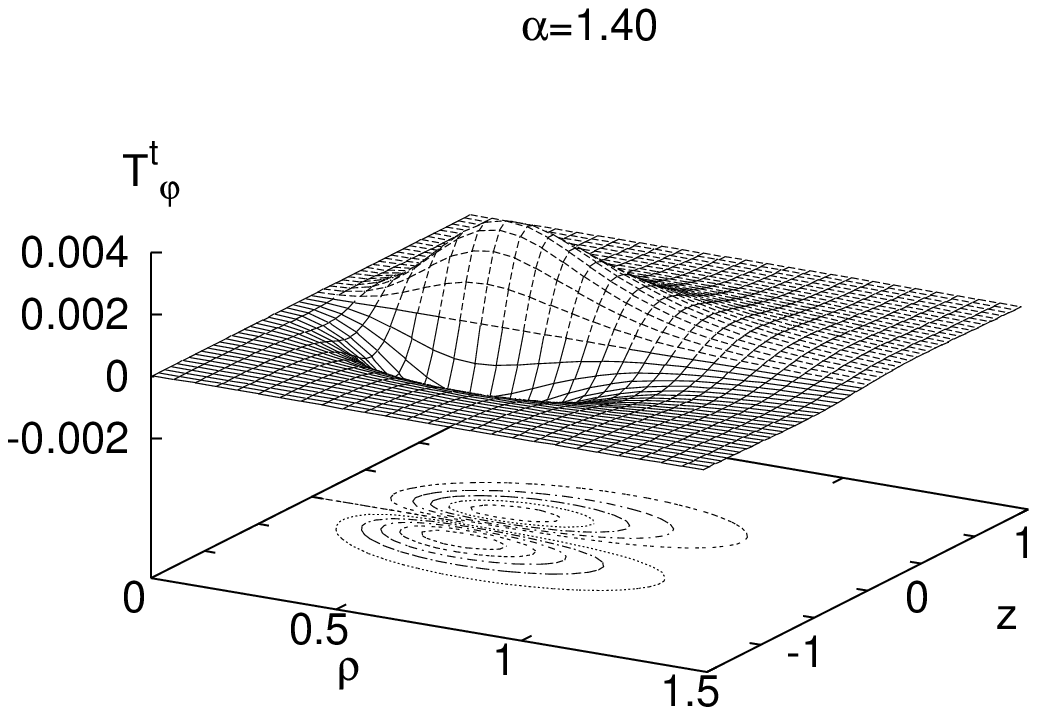} }
}\vspace{0.5cm}
{\bf Fig.~2} \small
The energy density $T^t_{\,t}$ (a) 
and the angular momentum density $T^t_{\,\varphi}$ (b)
are shown for a dyon solution with $n=2$, $\alpha=1.4$,
$\hat \gamma=0.32$.
\vspace{0.5cm}
}

\boldmath
\subsection{Rotating monopole-antimonopole pairs: $n=1$}
\unboldmath

Let us now consider rotating monopole-antimonopole pairs composed of
poles of charge $\pm 1$.
For these one expects an analogous
coupling constant dependence as for static monopole-antimonopole pairs,
i.e., when gravity is coupled, a branch of 
gravitating monopole-antimonopole pair solutions
emerges from the flat space solution,
and merges with a second branch of monopole-antimonopole pair solutions
at a maximal value of the coupling constant $\alpha_{\rm max}$.
The second branch then extends all the way back to $\alpha = 0$.
Along the second branch,
with decreasing $\alpha$, the solutions shrink to zero size \cite{MAP}.
Scaling the coordinates and the Higgs field with $\alpha$
however, leads to a limiting solution with finite size and 
finite scaled mass $\hat{M}$ \cite{MAP},
representing the lowest BM solution \cite{BM} of EYM theory.

The dependence of the rotating monopole-antimonopole pair solutions
on the coupling constant $\alpha$ has in part
been studied before \cite{tigran}.
There indeed two branches of solutions were found, 
however, these were not smoothly connected:
at a value $\alpha_{\rm cr}$ the mass of the solutions on both branches
agreed, but their angular momenta did not.
Thus the existence of further branches was hypothesized.

Repeating the numerical study reveals, however,
that both branches can be extended beyond $\alpha_{\rm cr}$,
up to a maximal value $\alpha_{\rm max}$,
where they merge.
Thus rotating monopole-antimonopole pair solutions indeed show the
expected coupling constant dependence,
except that the two branches of solutions cross before they merge.
This is illustrated in Figs.1. 
In \cite{tigran} the crossing point was interpreted as a
critical point $\alpha_{\rm cr}$.
The angular momentum of the solutions satisfies
the relation $J=Q$, since $n=1$. 
The magnetic moment of these solutions is exhibited in Fig.3.

Along the two branches,
the two nodes of the Higgs field,
which represent the locations of the magnetic poles,
move continuously closer together,
until they merge at the origin in the EYM limit on the second branch.
Thus the nodes of
the rotating monopole-antimonopole pair solutions also
exhibit the same $\alpha$ dependence as the nodes of static solutions 
\cite{MAP,KKS4}.

\noindent\parbox{\textwidth}{
\centerline{
(a)\mbox{\epsfysize=5.0cm \epsffile{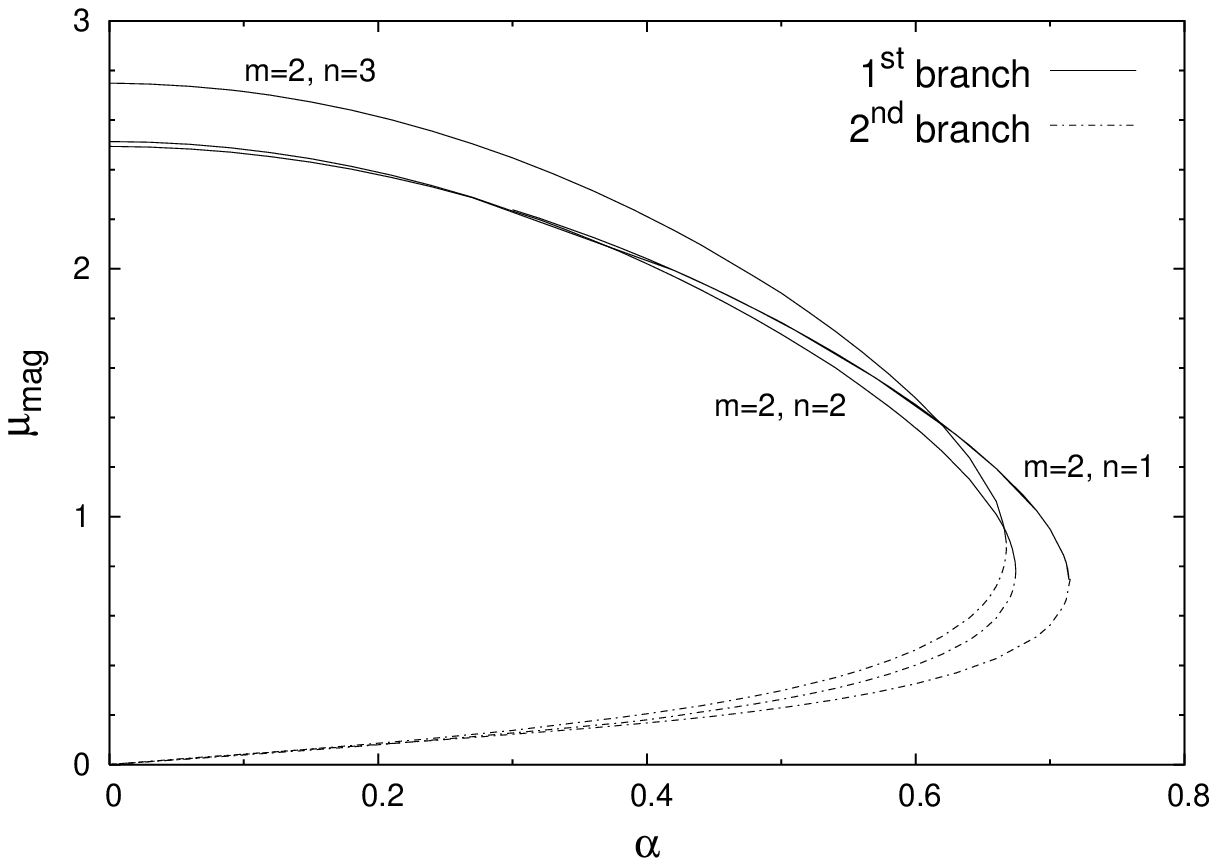} }
\hspace{2cm}
(b)\mbox{\epsfysize=5.0cm \epsffile{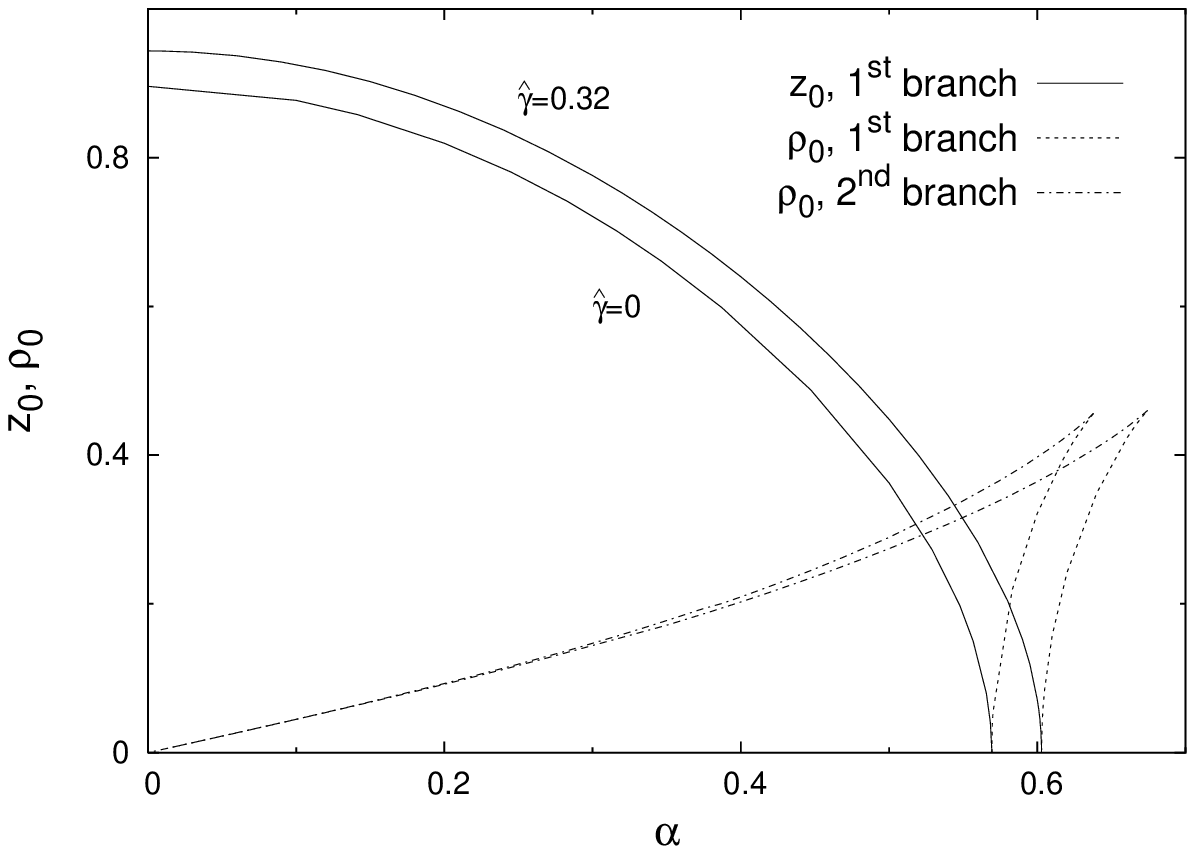} }
}\vspace{0.5cm}
{\bf Fig.~3} \small
The magnetic moment $\mu_{\rm mag}$ 
for monopole-antimonopole resp.~vortex ring solutions
with $n=1,2,3$ at $\hat \gamma=0.32$ (a) 
and the location of the nodes 
for monopole-antimonopole resp.~vortex ring solutions
with $n=2$ at $\hat \gamma=0.32$ and $\hat \gamma=0$ (b)
are shown as functions of the coupling constant $\alpha$.
\vspace{0.5cm}
}

\boldmath
\subsection{Rotating pairs and vortex rings: $n\ge 2$}
\unboldmath

For static monopole-antimonopole pairs 
composed of monopoles and antimonopoles of
charge $\pm 2$ \cite{KKS,tigran}, 
the $\alpha$ dependence is completely analogous as for pairs with
$n=1$.
Two branches of solutions exist, which merge at a maximal value
$\alpha_{\rm max}$.
The $\alpha$-dependence of the two nodes of the solutions was
not considered before \cite{KKS4}.

For rotating monopole-antimonopole pairs we also obtain two
branches of solutions. 
Again the two branches cross before they merge at a maximal value
$\alpha_{\rm max}$.
However, when the two nodes of the Higgs field are inspected, 
one finds a surprise.
The nodes merge at the origin at a value $\alpha_{\rm vr}$.
There the solutions change their character,
turning into vortex rings beyond $\alpha_{\rm vr}$,
since their Higgs field then vanishes on a ring in the $xy$-plane.
The nodal ring first increases in size, reaches a
maximum at $\alpha_{\rm max}$,
and then decreases to zero size in the EYM limit on the second branch.
The locations of the nodes and the nodal rings are exhibited in Fig.3.
As illustrated, the nodes of the static solutions possess an analogous
$\alpha$-dependence. 

For $\alpha < \alpha_{\rm vr}$, the energy density of the solutions
on the first branch consists of two tori, 
located symmetrically with respect to the $xy$-plane,
whereas for $\alpha > \alpha_{\rm vr}$, the energy density is a single torus,
as illustrated in Fig.~4.
The figure also shows that
the angular momentum density of the respective solutions is always
symmetrical with respect to the $xy$-plane.

\noindent\parbox{\textwidth}{
\centerline{
(a)\mbox{\epsfysize=6.0cm \epsffile{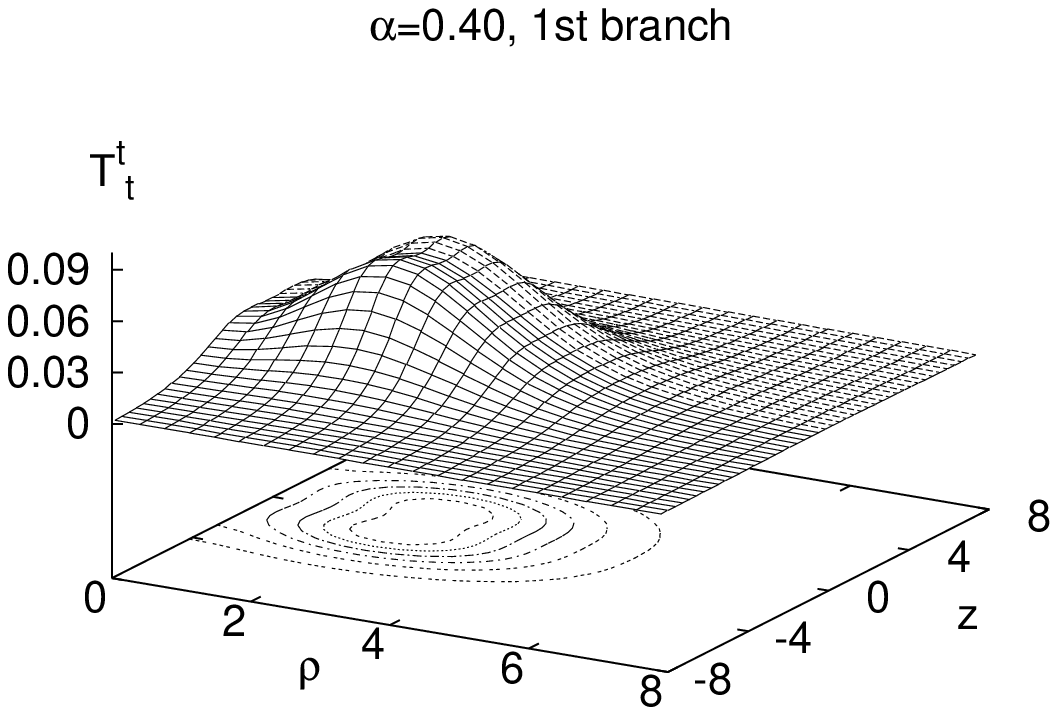} }
(b)\mbox{\epsfysize=6.0cm \epsffile{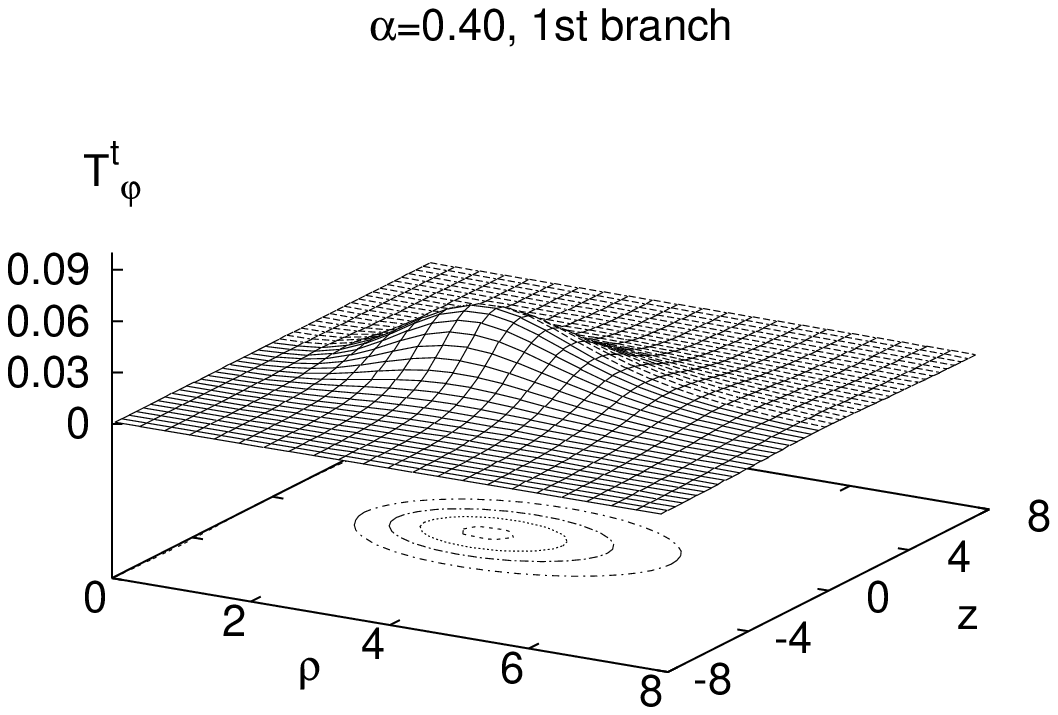} }
}\vspace{0.5cm}
\centerline{
(c)\mbox{\epsfysize=6.0cm \epsffile{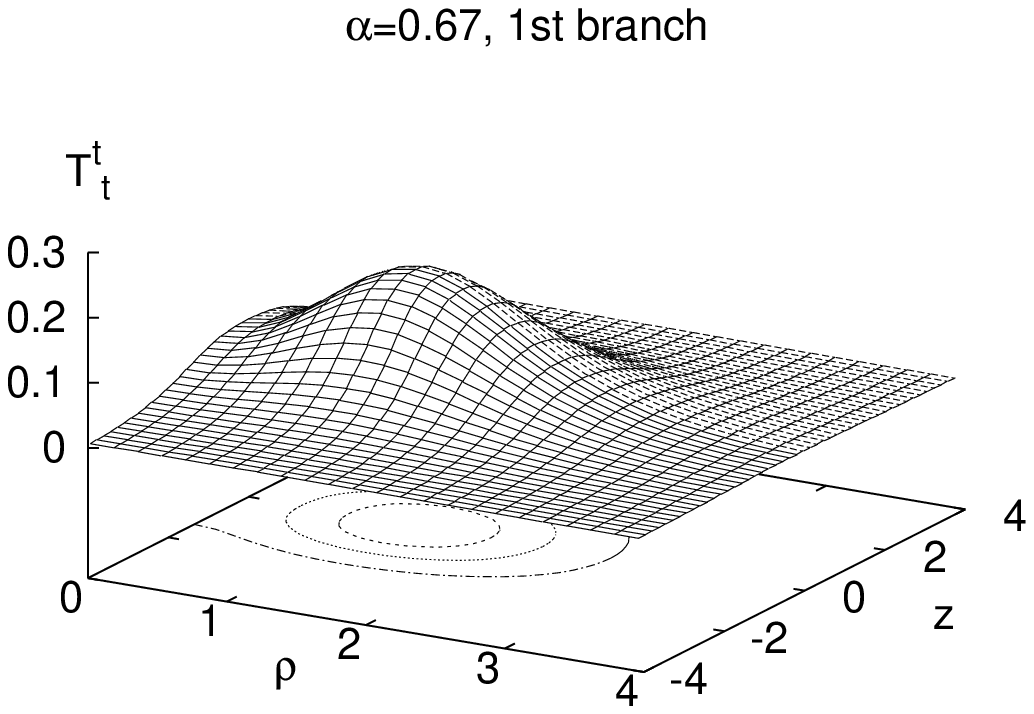} }
(d)\mbox{\epsfysize=6.0cm \epsffile{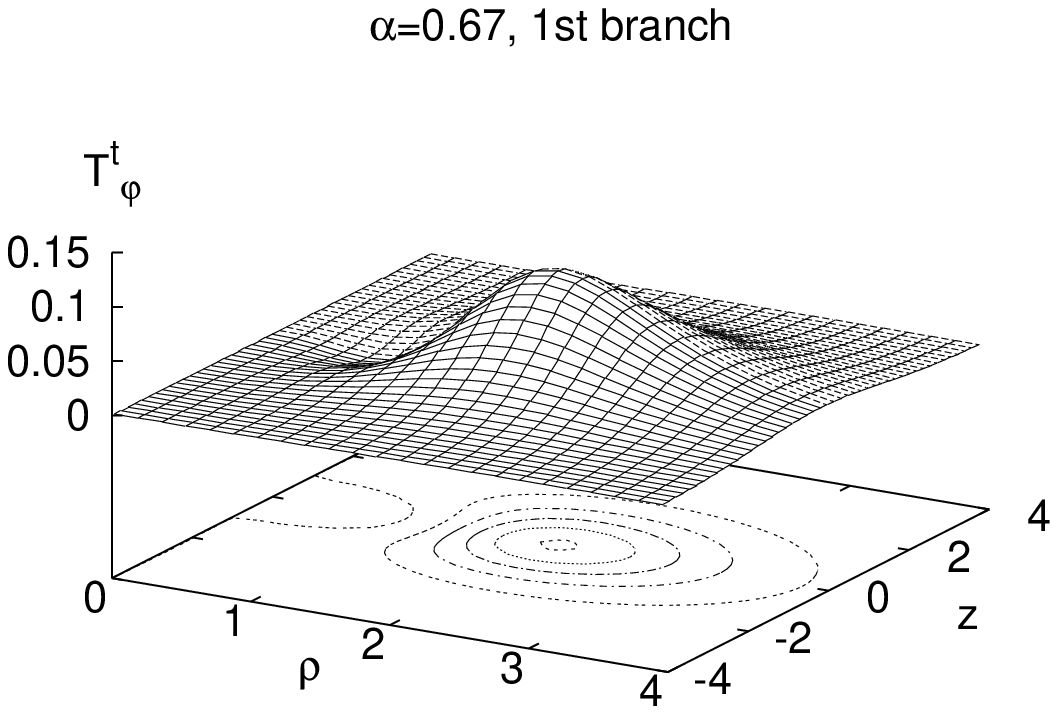} }
}\vspace{0.5cm}
{\bf Fig.~4} \small
The energy density $T^t_{\,t}$ (a,c)
and the angular momentum density $T^t_{\,\varphi}$ (b,d)
are shown for a monopole-antimonopole pair solution with $n=2$, $\alpha=0.4$,
$\hat \gamma=0.32$ (a,b)
and a vortex ring solution with $n=2$, $\alpha=0.67$, $\hat \gamma=0.32$ (c,d).
\vspace{0.5cm}
}

When $n=3$, already the static solutions correspond to vortex ring solutions
\cite{KKS,KKS4}. Like these,
the rotating vortex ring solutions possess two branches, merging
at $\alpha_{\rm max}$. 
As seen in Figs.~1 and 3,
the mass, charge and magnetic moment 
exhibit an analogous $\alpha$ dependence 
for these rotating vortex ring solutions
as for the rotating $n=1$ and $n=2$ solutions.
The nodal ring continuously decreases in size along both branches,
shrinking to zero size in the EYM limit.

\section{Conclusions}

We have constructed stationary dyons and rotating monopole-antimonopole
pairs and vortex rings in EYMH theory.
Gravitating dyons cannot rotate, since they carry magnetic charge \cite{radu}.
We have shown, that gravitating
dyons with magnetic charge $n \ge 2$, 
however, are not static but stationary,
possessing finite angular momentum density and a non-diagonal metric
(see also \cite{BHR}).
Electrically charged monopole-antimonopole pairs 
as well as (magnetically neutral) vortex rings,
on the other hand, must rotate with angular momentum $J=nQ$.
For monopole-antimonopole pairs,
consisting of singly charged magnetic poles,
we have resolved the critical behaviour,
by showing that the two branches of solutions cross before they merge.
The same feature is present for rotating monopole-antimonopole pairs
with $n\ge 2$ and rotating vortex rings.
Interestingly,
for monopole-antimonopole pairs, consisting of doubly charged magnetic poles,
we have observed a transition to vortex rings at a value $\alpha_{\rm vr}$
on the first branch. The transition value depends on 
the amount of electric charge present and thus on the rotation.

For the dyons, monopole-antimonopole pair solutions and vortex rings 
presented here,
we have restricted the integers $m$ and $n$ in the 
general ansatz for the matter fields to $m=1-2$ and $n=1-3$.
The solutions thus represent only the simplest types of EYMH solutions.
For larger values of $m$ and $n$ new types of static solutions appear
\cite{KKS,KKS4,IKKS}, representing e.g.~monopole-antimonopole chains
and multi-vortex solutions.
Study of their stationary or rotating generalizations may lead to further
surprises.

\begin{acknowledgments}
B.K. gratefully acknowledges support by the DFG under contract
KU612/9-1.
\end{acknowledgments}

\newpage


\begin{thebibliography}{000}
\bibitem{foot}
 For gravitating monopoles
 a second branch is present only for small and vanishing
 Higgs self-coupling constant \cite{gmono}.
\end{thebibliography}
\end{document}